\documentclass[aps,prd,twocolumn,floatfix,nofootinbib,nogroupedaddress,showpacs]{revtex4}
\usepackage{dcolumn}
\usepackage{graphicx}

\newcommand{\be}{\begin{eqnarray}}
\newcommand{\ee}{\end{eqnarray}}
\newcommand\del{\partial}
\newcommand{\nn}{\nonumber}
\newcommand{\mui}{\mu_{\rm iso}}
\newcommand{\hatmui}{\hat{\mu}_{\rm iso}}
\newcommand{\mat}{\left ( \begin{array}{cc}}
\newcommand{\emat}{\end{array} \right )}
\newcommand{\Str} {{\rm Str}\,}
\def\bml{\begin{mathletters}}
\def\eml{\end{mathletters}}
\def\ba{\begin{array}}
\def\ea{\end{array}}

\def\nn{\nonumber}
\def\to{\rightarrow}
\def\la{\lambda}

\def\ep{\epsilon}

\def\tr{{\rm tr}\,}
\def\Tr{{\rm Tr}\,}
\newlength{\bredde}
\def\slash#1{\settowidth{\bredde}{$#1$}\ifmmode\,\raisebox{.15ex}{/}
\hspace*{-\bredde} #1\else$\,\raisebox{.15ex}{/}\hspace*{-\bredde} #1$\fi}
\def\Sl#1{\rlap{\raisebox{.15ex}{$\mskip 4 mu /$}}#1}  
\begin{document}
\title{Microscopic eigenvalue correlations in QCD with imaginary isospin chemical potential}
\author{P. H. Damgaard}
\affiliation{
The Niels Bohr Institute, Blegdamsvej 17, DK-2100 Copenhagen \O,
Denmark
}
\author{U. M. Heller}
\affiliation{
American Physical Society, One Research Road, Box 9000, Ridge,
NY 11961-9000, USA
}
\author{K. Splittorff}
\affiliation{
The Niels~Bohr~Institute, Blegdamsvej 17, DK-2100 Copenhagen \O, Denmark
}
\author{B. Svetitsky}
\affiliation{School of Physics and Astronomy, Raymond and Beverly Sackler
Faculty of Exact Sciences, Tel Aviv University, 69978 Tel Aviv, Israel
}
\author{D.~Toublan}
\affiliation {Physics Department, University of Maryland, College
Park, MD 20742
}

\date{\today}
\begin{abstract}
We consider the chiral limit of QCD subjected to an imaginary isospin
chemical potential. In the $\epsilon$-regime of the theory we can
perform precise analytical calculations based on the zero-momentum Goldstone
modes in the low-energy effective theory. We present results
for the spectral correlation functions of the associated Dirac
operators.
\end{abstract}
\maketitle

\section{Introduction}

An isospin chemical potential provides a way to ``twist'' the usual Dirac
operator in two different directions.
A real isospin chemical potential \cite{AKW} gives a fermion determinant that is real and positive, and thus
amenable to numerical simulations; it does not, however, preserve anti-hermiticity of the Dirac operator itself.  An imaginary value of the isospin chemical potential, on the other hand, gives massless Dirac operators that are anti-hermitian; their eigenvalues thus 
lie on the imaginary axis instead of spreading out into the complex plane. This makes
imaginary isospin chemical potential a useful parameter for
deformation of the Dirac eigenvalue spectrum. 

Recently we have noted \cite{DHSS,DHSST} that a particular spectral two-point correlation
function of Dirac operator eigenvalues near the origin yields a direct
way of determining the pion decay constant, $F_{\pi}$, from lattice gauge
theory simulations. An alternative proposal, also using imaginary isospin chemical potential, is to use the
distortion of the mass-dependent chiral condensate%
\footnote{The twisted boundary conditions used there are equivalent to
an imaginary isospin chemical potential.}
\cite{MT}.
For the quenched theory it has also been shown that one can use an ordinary
(real) baryon chemical potential to extract the pion decay constant from
lattice distributions of the Dirac eigenvalues \cite{OW,AB}. For gauge group
SU(3) and quarks in the fundamental representation, the
Dirac operator spectrum with real baryon chemical potential is
complex, and the theory with dynamical 
quarks is difficult to simulate directly due to a complex fermion determinant.
For this reason imaginary isospin chemical potential is a more convenient
strategy.

For a given non-Abelian
gauge potential $A_\mu(x)$ we study the two Dirac operators
\be
D_+\psi_+^{(n)} \equiv [\Sl{D}(A)+i\mui\gamma_0]\psi_+^{(n)}=
i\lambda_+^{(n)}\psi_+^{(n)} \label{D+}
\ee
and
\be
D_-\psi_-^{(n)} \equiv [\Sl{D}(A)-i\mui\gamma_0]\psi_-^{(n)}
= i\lambda_-^{(n)}\psi_-^{(n)}, \label{D-}
\ee
where $\mui$ is real.
Both operators $D_{\pm}$ are anti-hermitian, and the eigenvalues
$\lambda_\pm^{(n)}$ therefore lie on the real line. An imaginary
isospin chemical potential can be viewed as an external constant Abelian gauge
potential $A_0$ that couples to the $u$ and $d$ quark with opposite
charges.

Much work has gone into understanding gauge theories
at real isospin chemical potential 
\cite{SoS,KSS,KT,STV,KoS,Loewe:2002tw,Barducci:2003un,Barducci:2004tt}, often
in terms of the
effective low-energy theory. Here we consider the effective theory, a
chiral Lagrangian, in the presence of an imaginary isospin chemical potential.
With the usual pattern of spontaneous chiral symmetry breaking for two
light flavors, the theory is described by a Lie-group valued field
$U(x) \in $~SU(2).

Our focus will be on the
so-called $\epsilon$-regime of QCD, where the chiral Lagrangian is treated
as a perturbative expansion around the zero-momentum modes in a finite volume
$V$ \cite{LS}. Roughly speaking, we are dealing with an expansion
in $1/L$ (where $L$ is the linear extent of the finite volume) rather than
the usual expansion in a small momentum $p$. There is a well known and intriguing connection between this regime and a universal limit of Random Matrix
Theory \cite{Jac,ADMN,OTV,DOTV,TVbeta,TV1,TV},
but here we will stay entirely within the framework of the effective chiral
Lagrangians.

When the chemical potential is included, the power counting of the
$\epsilon$-expansion must be reconsidered. One factorizes the field
as $U(x) = U_0\exp[i\sqrt{2}\phi(x)/F_{\pi}]$, where $U_0$ is the zero
momentum part that will be treated exactly, and $\phi(x)$ represents the
fluctuation fields (without zero modes).
It turns out that the
naive guess provides a consistent counting: To leading order one keeps
only the static modes in the path integral, while the fluctuation degrees
of freedom decouple. This is not completely obvious at first glance,
but can be seen as
follows. The coupling to the imaginary isospin chemical potential in
the chiral Lagrangian is dictated by the way it couples at the quark level [Eqs.~(\ref{D+}) and~(\ref{D-})]. Vector sources of that
kind give rise to a covariant time derivative in the effective
SU(2) Lagrangian \cite{KST,SoS},
\be
\partial_0U(x) ~\to~ \nabla_0U(x) = \partial_0U(x) -i \mui[\sigma_3,U(x)] ,
\label{Ucov}
\ee
where $\sigma_3$ is the usual Pauli matrix.
The leading-order terms in the effective Lagrangian then read
\be
{\cal L} &=& \frac{F_{\pi}}{4}\Tr\left[\nabla_0U(x)\nabla_0U^\dagger(x) +
\partial_iU(x)\partial_iU^\dagger(x)\right] \nonumber\\ &&-
\frac{\Sigma}{2}\Tr\left[{\cal M}U^{\dagger}(x)
+ {\cal M}^{\dagger}U(x)\right], \label{L2}
\ee
where ${\cal M}={\rm diag}(m_u,m_d)$ is the quark mass matrix and $\Sigma$ 
is the chiral condensate. When we expand
\be
U(x) = U_0\left[1 + i\sqrt{2}\phi(x)/F_{\pi} + \cdots\right] \label{Ueps},
\ee
this produces the usual kinetic term for $\phi(x)$.
Let us recall the power counting in the $\epsilon$-expansion \cite{LS}:
We assume $m_{\pi} \sim p^2 = {\cal O}(\epsilon^2)$ while $\phi(x) \sim 1/L
= {\cal O}(\epsilon)$, and a consistent power counting for the $\mui$-term is $\mui =
{\cal O}(\epsilon^2)$. Indeed, when we expand the covariant derivative
(\ref{Ucov}) using Eq.~(\ref{Ueps}) the leading contribution becomes
\be
\nabla_0U(x) = i\sqrt{2}/F_{\pi} \partial_0\phi(x) -i \mui[\sigma_3,U_0] + \cdots.
\label{nablaleading}
\ee
In the chiral Lagrangian (\ref{L2}), the mixed terms $\partial_0\phi(x)[\sigma_3,U_0]$
produce only boundary contributions and play no role here.
Thus, to leading order in the $\epsilon$-expansion the fluctuation field
$\phi(x)$ gives rise only to the kinetic energy term
$$
\int\! d^4x\, \frac{1}{2}\Tr\partial_{\mu}\phi(x)\partial_{\mu}\phi(x),
$$
which decouples as in the theory with $\mui = 0$.

Collecting the remaining terms
we see that the leading contribution to the partition function in the $\epsilon$-regime
is the zero-dimensional integral
\begin{widetext}
\be
{\cal Z}({\cal M};i\mui) =
\int_{SU(2)} dU\, e^{\frac{1}{4}VF_\pi^2\mui^2
\Tr [U,\sigma_3][U^\dagger,\sigma_3] 
+ \frac 12 \Sigma V
\Tr({\cal M}^\dagger U + {\cal M}U^\dagger)} ,
\label{zeffQCD}
\ee
where we have dropped the $0$-suffix on the group element $U \in$~SU(2). Projection
onto fixed gauge field topology $\nu$ \cite{LS} is done by a Fourier transform, and
amounts to the simple modification
\be
{\cal Z}^{\nu}({\cal M};i\mui) =
\int_{U(2)} dU \,(\det U)^{\nu}e^{\frac{1}{4}VF_\pi^2\mui^2
\Tr [U,\sigma_3][U^\dagger,\sigma_3] + \frac 12 \Sigma V
\Tr({\cal M}^\dagger U + {\cal M}U^\dagger)} ~.
\label{zeffnu}
\ee
\end{widetext}
One sees that the leading-order contribution to the $\epsilon$-regime depends
only on the scaling variables $\hat{m}_i\equiv m_i\Sigma V$ (where $m_i$ are the quark masses)
and $\hatmui^2\equiv\mui^2F_{\pi}^2V$. Both of these scaling variables are of order 1 in the
$\epsilon$-counting.

Effective partition functions related to (\ref{zeffnu}), and to its generalizations to more quark
flavors of both kinds of statistics, have been studied in great detail recently \cite{TV,SV,O,AOSV,MT,AFV},
and much has been learned about them. A particularly important feature for what follows
is that such partition functions satisfy a series of exact relationships relating theories
with different numbers of quark flavors to each other. The origin of the formalism lies in the theory
of certain integrable systems, but we need here only the identities themselves,
which are known under the names of Painlev\'{e} and Toda lattice equations
\cite{K,SV0}. These equations will be used to provide a non-perturbative definition
of a {\em replica limit}, which in turn is needed to compute spectral correlation functions
of the Dirac operator eigenvalues.

We have organized this paper as follows. In Sec.~II we reconsider the case of quenched QCD,
for which a comparison with lattice gauge theory simulations has already been
presented \cite{DHSS}. In that paper the analytical results
were stated without proof; here we provide the details. The main idea is to focus on a mixed
spectral correlation function which is extremely sensitive to imaginary
isospin chemical potential. In Sec.~III we turn to the physically interesting case
of two light quark flavors. The same two-point spectral correlation function is far more
difficult to determine analytically. In a previous paper \cite{DHSST} we briefly
reported the final results, and showed how well they compare with lattice gauge
theory simulations. The bulk of this paper, including all of Sec.~III,
is dedicated to the detailed derivation of just those results. Finally, Sec.~IV contains
our conclusions and an outlook on future work.

\section{Quenched theory\label{sec2}}

In order to consider the quenched analogue of the situation outlined in the Introduction
we need to define the quenched limit on the effective field theory side. This issue
was first resolved in Ref.~\cite{BG} by means of a chiral Lagrangian living on a
graded (``supersymmetric'') coset of spontaneous chiral symmetry breaking. An alternative,
closer to the approach we shall pursue in this paper, relies on the replica method
\cite{DS}. We stress already here that ``quenching,'' be it by means of replicas or
quark partners of bosonic statistics, is required even in the case of dynamical quarks
if one wishes to compute spectral correlation functions of the Dirac
operator.  Indeed, these methods are the only known approaches that allow
access to the low-energy Dirac spectrum from effective field theory.

The result of the quenched calculation was briefly stated
in Ref.~\cite{DHSS}, which otherwise focused on the high numerical precision
that can be reached for $F_{\pi}$ with the proposed method for
measuring it. As a warm-up exercise for the $N_f=2$ calculation
we will here give the main ingredients behind this quenched result.  We
stress that the steps we follow are the same for both the quenched and
dynamical cases; the only difference is that each step is simpler in the
quenched case. We first define a two-point correlation function which is very
sensitive to $F_\pi$. This correlation function can be obtained from a
susceptibility that we define and calculate in the effective theory in the
$\epsilon$-regime. This last calculation, performed here using the replica
method, is the most difficult part.  It requires the use of  generating
functions that are explicitly derived.

\begin{figure}[htb]
\begin{center}
\includegraphics[width=.95\columnwidth]{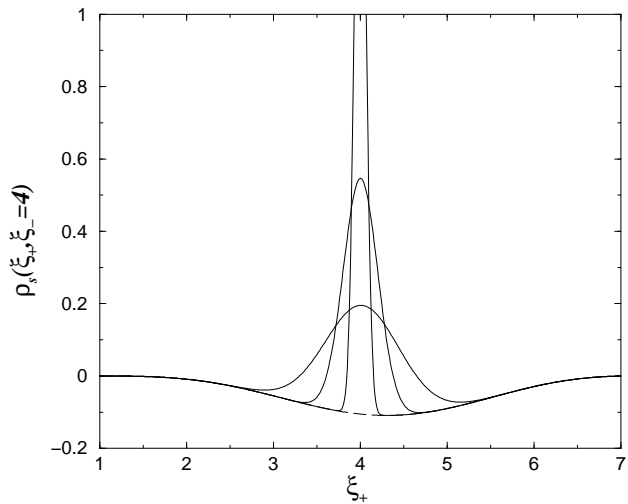}
\caption{\label{fig:1}
The quenched two point-correlation function (\ref{rho2Qmui}) with one
eigenvalue arbitrarily fixed at $\xi_- = 4$ as a function of the other
eigenvalue $\xi_+$. The full curves correspond to $\hatmui^2 =0.001$,
$0.01$ and $0.05$. The curve with the narrowest peak corresponds to
the smallest value of $\hatmui$. Also shown (dashed) is the curve
for $\hatmui=0$ where the peak becomes a $\delta$-function at 
$\xi_-=\xi_+=4$. The $\delta$-function has been suppressed from this figure.}
\end{center}
\end{figure}

The method of Ref.~\cite{DHSS} is to consider the ``mixed'' two-point
spectral correlation function of the Dirac operators $D_{\pm}$ defined
in Eqs.~(\ref{D+}) and~(\ref{D-}),
\begin{widetext}
\be
\rho(\lambda_+,\lambda_-;i\mui) \equiv  \left\langle
\sum_n \delta\left(\lambda_+ - \lambda_+^{(n)}\right)\sum_m
 \delta\left(\lambda_- -\lambda_-^{(m)}\right)\right\rangle 
     - \left\langle \sum_n \delta\left(\lambda_+ -\lambda_+^{(n)}\right)\right\rangle
         \left\langle\sum_m \delta\left(\lambda_- - \lambda_-^{(m)}\right)\right\rangle ,
\label{rho2Q}
\ee
where the averages are performed over
the pure Yang-Mills partition function.
In order to reach the $\epsilon$-regime, this correlator is considered in the
microscopic limit
\be
\rho_s(\xi_+,\xi_-;i\hatmui)
\equiv \lim_{V\to\infty}\frac{1}{\Sigma^2  V^2}
\rho\left(\frac{\xi_+}{\Sigma V},\frac{\xi_-}{\Sigma V};\frac{i\hatmui}
{F_\pi\sqrt{V}}\right).
\ee

A generating function for the spectral correlation function (\ref{rho2Q}) is the
mixed scalar susceptibility,
\be
\chi(m_+,m_-;i\mui) \equiv 
\left\langle \tr \frac{1}{D_+ + m_+}\tr
 \frac{1}{D_- + m_-}\right\rangle  
     - \left\langle \tr \frac{1}{D_+ + m_+}\right\rangle
         \left\langle\tr\frac{1}{D_- + m_-}\right\rangle.
\label{chiQ}
\ee
As above, the averages are performed over the
pure Yang-Mills partition function. (Note that this partition function
is independent of $m_+$ and
$m_-$.) Written in an eigenvalue representation, Eq.~(\ref{chiQ})
becomes
\be
\chi(m_+,m_-;i\mui)  = \left\langle \sum_n \frac{1}{i\lambda_+^{(n)}+m_+}\sum_m
 \frac{1}{i\lambda_-^{(m)} + m_-}\right\rangle  
   - \left\langle \sum_n \frac{1}{i\lambda_+^{(n)}+m_+}\right\rangle
         \left\langle\sum_m\frac{1}{i\lambda_-^{(m)} + m_-}\right\rangle.
\ee
If one knows this function analytically,
the spectral two-point function (\ref{rho2Q})
can be computed from the discontinuity across the imaginary axis \cite{DOTV,TV1},
\be
\rho(\lambda_+,\lambda_-;i\mui) & = & \frac1{4\pi^2}
{\rm  Disc}[\chi(m_+,m_-;\mui)]_{m_+=i\lambda_+,m_-=i\lambda_-} \nn\\
& = & \frac1{4\pi^2}\lim_{\ep\to0^+}[
 \chi(i\lambda_+ +\ep,i\lambda_- +\ep;\mui)
-\chi(i\lambda_+-\ep,i\lambda_-+\ep;\mui) \nn\\
&& \qquad \qquad \quad -\chi(i\lambda_+ +\ep,i\lambda_- -\ep;\mui)
+\chi(i\lambda_+ -\ep,i\lambda_- -\ep;\mui)] , \label{chidisc}
\ee
which is the inverse of the relation
\be
\chi(m_+,m_-;i\mui) & = & \int_{-\infty}^\infty d\lambda_+\,d\lambda_-\,
\frac{\rho(\lambda_+,\lambda_-;i\mui)}{(i\lambda_++m_+)(i\lambda_- + m_-)} .
\ee

In the replica formalism the mixed scalar susceptibility (\ref{chiQ}) can
be defined as \cite{DS}
\be
\chi(m_+,m_-;i\mui)\equiv\lim_{n\to 0} \frac{1}{n^2}\del_{m_+}
\del_{m_-} \log {\cal Z}^{\nu}_{2n}(m_+,m_-;i\mui),
\label{chi-replica}
\ee
where ${\cal Z}^{\nu}_{2n}(m_+,m_-;i\mui)$ is the effective partition function of $2n$ replicated
quark flavors.
Half of these have degenerate masses $m_+$ and chemical potential
$i\mui$, while the remaining $n$ flavors have degenerate masses $m_-$ and chemical potential
$-i\mui$. At the level of the fundamental theory,
\be
{\cal Z}^{\nu}_{2n}(m_+,m_-;i\mui) = \int[{\rm d}A]_\nu
\det(D_+ + m_+)^n\det(D_- + m_-)^n
      e^{-S_{\rm YM}(A)} .
\label{Z2nQCD}
\ee
The leading contribution in the $\epsilon$-regime is analogous to the SU(2)
case discussed in the introduction, and reads \cite{SV}
\be
{\cal Z}^{\nu}_{2n}(\hat{m}_+,\hat{m}_-;i\hat\mui) = 
\int_{U \in U(2n)} dU\, \det(U)^\nu e^{\frac{1}{4}\hat\mui^2
\Tr [U,B][U^\dagger,B] + \frac 12 \Tr(M^\dagger U + MU^\dagger)},
\label{zeffrepQ}
\ee
where $B = {\rm diag}({\mathbf 1}_n,-{\mathbf 1}_n)$ and
$M = {\rm diag}(\hat{m}_+,\ldots,\hat{m}_+,\hat{m}_-,\ldots,\hat{m}_-)$, in
terms of the scaling variables
\be
\hat{m}_\pm \equiv m_\pm \Sigma V \ \ \ {\rm and} \ \ \  \hatmui \equiv \mui F_\pi \sqrt{V}.
\ee
We also use ${\cal Z}$ to denote the partition function in the $\epsilon$-regime, but to distinguish it from the general partition function (\ref{Z2nQCD}), we write it explicitly as a function of the scaled variables.
Because these partition functions
(\ref{zeffrepQ}) are linked to an integrable structure (of an otherwise unrelated
Hamiltonian system), it turns out that all can be derived from just the
$n=1$ case \cite{footnote},
\be
{\cal Z}^{\nu}_2(\hat{m}_+,\hat{m}_-;i\hatmui) =  e^{-2\hatmui^2}\int_0^1 dt \,t
e^{2\hatmui^2t^2} I_\nu(t \hat{m}_+)I_\nu (t \hat{m}_-),
\label{zeff2mui}
\ee
by means of \cite{SV}
\be
{\cal Z}^{\nu}_{2n}(\hat{m}_+,\hat{m}_-;i\hatmui) =
\frac{D_n}{(\hat{m}_+ \hat{m}_-)^{n(n-1)}} \det \left[(\hat{m}_+\del_{\hat{m}_+})^k
  (\hat{m}_-\del_{\hat{m}_-})^l {\cal Z}^{\nu}_2(\hat{m}_+,\hat{m}_-;i\hatmui)\right]_{k,l=0,1,\ldots,n-1}.
\label{zeff2nmui}
\ee
Here $D_n$ is a normalization factor whose exact value need not concern us here. It
is chosen so that the effective partition functions (\ref{zeffrepQ}) satisfy the Toda lattice
equation ($D_n$ determines the coefficient of proportionality) \cite{SV}
\be
\hat{m}_+\del_{\hat{m}_+} \hat{m}_-\del_{\hat{m}_-} \log {\cal Z}^{\nu}_{2n}(\hat{m}_+,\hat{m}_-;i\hatmui)
= 4n^2(\hat{m}_+\hat{m}_-)^2\frac
  {{\cal Z}^{\nu}_{2n+2}(\hat{m}_+,\hat{m}_-;i\hatmui){\cal Z}^\nu_{2n-2}(\hat{m}_+,\hat{m}_-;i\hatmui)}
{[{\cal Z}^{\nu}_{2n}(\hat{m}_+,\hat{m}_-;i\hatmui)]^2} ~.
\label{TodaQ}
\ee
Using this exact equation to define the replica limit, we obtain
\be
\chi(\hat{m}_+,\hat{m}_-;i\hatmui)= 4\hat{m}_+\hat{m}_- {\cal Z}^{\nu}_{2}(\hat{m}_+,\hat{m}_-;i\hatmui){\cal Z}^{\nu}_{-2}(\hat{m}_+,\hat{m}_-;i\hatmui) ~.
\label{suscepQ}
\ee

Apart from the U(2) partition function (\ref{zeffnu}), now given by (\ref{zeff2mui}), we also
need ${\cal Z}^\nu_{-2}(\hat{m}_+,\hat{m}_-;i\hatmui)$. This is the
partition function of $-2$ fermionic quarks, i.e., 
2 quarks of {\em bosonic statistics}.
Like its fermionic analogue, the effective bosonic partition function is determined by
the symmetries of the
underlying QCD Lagrangian, now with bosonic quarks. In addition,
in the bosonic case one must carefully enforce the convergence of the
partition function \cite{DOTV,SS}, a problem that is
absent in the fermionic case due to the nature of Grassmann integration. As the purely
imaginary chemical potential does not change the hermiticity properties
of the Dirac operator, this does not lead to additional constraints as it
does for real chemical potential \cite{SV,AOSV}. The bosonic partition function can be written as
the integral
\be
{\cal Z}^\nu_{-2}(\hat{m}_+,\hat{m}_-;i\hatmui) =
\int \frac{dQ\,\theta(Q)}{{\det}^2 Q}  \det(Q)^\nu
e^{-\frac{1}{4}\hatmui^2 \tr[Q,B][Q^{-1},B]
-\frac{1}{2}\tr({\cal M}^\dagger Q+Q^{-1}{\cal M})},
\label{zeffbosQ}
\ee
where $dQ\,\theta(Q)/{\det}^2Q$ is the integration measure on positive definite Hermitian matrices. Using the parametrization
\be
Q = e^t \mat e^r \cosh s  & e^{i\theta}\sinh s \\
          e^{-i\theta}\sinh s &   e^{-r} \cosh s \emat,
\ee
where
\be
r \in [ -\infty, \infty ], \quad
s \in [ -\infty, \infty ], \quad
t \in [ -\infty, \infty ], \quad
\theta \in [ 0, \pi ],
\ee
we find
\be
{\cal Z}^\nu_{-2}(\hat{m}_+,\hat{m}_-;i\hatmui) = e^{2\hatmui^2}\int_1^\infty dt\, t
e^{-2\hatmui^2t^2} K_\nu(t \hat{m}_+)K_\nu (t \hat{m}_-) .
\label{Z-1mui}
\ee
At $\hatmui=0$ this integral can be done analytically,
\be
{\cal Z}^\nu_{-2}(\hat{m}_+,\hat{m}_-;i\hatmui=0) =
\frac{\hat{m}_+K_{\nu+1}(\hat{m}_+)K_\nu(\hat{m}_-)
- \hat{m}_-K_{\nu+1}(\hat{m}_+)K_\nu(\hat{m}_-)}
{\hat{m}_+^2-\hat{m}_-^2} ,
\label{Z-1mu0}
\ee
and the result agrees with the expression at $\mui=0$ that was derived in Refs.~\cite{DV,FA}. The main difference between this and the corresponding fermionic result
is the replacement of modified Bessel functions $I_n(x)$ by $K_n(x)$. This can be traced
back to the non-compact integration range in (\ref{zeffbosQ}), which in turn follows from
the symmetries and convergence requirements of the theory with bosonic quarks.

With the above ingredients we immediately find the mixed scalar susceptibility
from Eq.~(\ref{suscepQ}),
\be
\chi(\hat{m}_+,\hat{m}_-;i\hatmui)  = 4\hat{m}_+\hat{m}_-\left[\int_0^1 dt \,t
e^{2\hatmui^2t^2} I_\nu(t \hat{m}_+)I_\nu (t
\hat{m}_-) \right]\left[
\int_1^\infty dt \,t
e^{-2\hatmui^2t^2} K_\nu(t \hat{m}_+)K_\nu (t \hat{m}_-) \right].
\label{suscepQexpl}
\ee
Taking the discontinuity as dictated by Eq.~(\ref{chidisc}), we finally obtain
the desired quenched spectral correlation function,
\be
\label{rho2Qmui}
\rho_s(\xi_+,\xi_-;i\hatmui) & = & \xi_+\xi_- \left[\int_0^1 dt\, t
e^{2\hatmui^2t^2} J_\nu(t \xi_+)J_\nu (t \xi_-)\right]\left[
\int_1^\infty dt\,t e^{-2\hatmui^2t^2}
J_\nu(t \xi_+)J_\nu (t \xi_-)\right]  \\
&=& \xi_+\xi_- \left[\int_0^1 dt \,t
e^{2\hatmui^2t^2} J_\nu(t \xi_+)J_\nu (t \xi_-)\right] \nn\\
&&\times \left[\frac{1}{4 \hatmui^2}
\exp\left(-\frac{(\xi_+^2 + \xi_-^2)}{8 \hatmui^2}\right)
I_\nu\left(\frac{\xi_+\xi_-}{4 \hatmui^2}\right)
-\int_0^1 dt \,t e^{-2\hatmui^2t^2}
J_\nu(t \xi_+)J_\nu (t \xi_-)\right]. \nn
\ee
In the last line we have traded one non-compact integral for an integral over the compact interval $[0,1]$. This is convenient if one wishes
to evaluate the expression numerically. 

Again it is useful to check the limiting
case $\mui=0$, where the above expression can be simplified. Using the orthogonality
properties of modified Bessel functions on the interval $[0,\infty]$, we have
\be
\label{idBessel}
 \int_1^\infty dt \,t J_\nu(t \xi_+)J_\nu (t \xi_-)
& = & \int_0^\infty dt \,t J_\nu(t \xi_+)J_\nu (t \xi_-)-
 \int_0^1      dt \,t J_\nu(t \xi_+)J_\nu (t \xi_+) \nn\\[2pt]
&=& \frac{1}{\xi_+}\delta(\xi_+ - \xi_-)-
 \frac{\xi_+ J_{\nu+1}(\xi_+)J_{\nu}(\xi_-)-\xi_- J_{\nu+1}(\xi_-)J_{\nu}(\xi_+)}
{(\xi_+^2 - \xi_-^2)} ~.
\ee

This allows us to rewrite the above expression for $\mui=0$ as
\be
\rho_s(\xi_+,\xi_-;i\hatmui=0) & = &
\delta(\xi_+ - \xi_-)\frac{\xi_+}{2}[J_\nu^2(\xi_+)
-J_{\nu+1}(\xi_+)J_{\nu-1}(\xi_+)] \cr
&& -\frac{\xi_+\xi_-}{(\xi_+^2 - \xi_-^2)^2}
\left[\xi_+ J_{\nu+1}(\xi_+)J_{\nu}(\xi_-)
    - \xi_- J_{\nu+1}(\xi_-)J_{\nu}(\xi_+)\right]^2.\label{rho2Qnomui}
\ee
\end{widetext}
which agrees with the known result \cite{Jac, ADMN,TV1}. When the two eigenvalues
$\xi_\pm$ coincide there is an explicit $\delta$-function contribution whose
coefficient is given by the spectral one-point function. This is due to
eigenvalue auto-correlation. 
The way that the finite imaginary isospin chemical potential $\mui$ ``resolves'' this
$\delta$-function contribution (because the eigenvalues $\xi_-$ and $\xi_+$
are now associated with two different Dirac operators) is quite spectacular. We show
in Fig.~\ref{fig:1} 
the quenched spectral two-point function (\ref{rho2Qmui}) with one
eigenvalue arbitrarily 
fixed at $\xi_- = 4$ as a function of the other eigenvalue $\xi_+$
for $\hatmui^2=0$, 0.001, 0.01, and 0.05. The pronounced peak around
$\xi_+=4$ is precisely the remnant of the $\delta$-function at $\xi_- = \xi_+$.
It has been shown in Ref.~\cite{DHSS} how this spreading out of the $\delta$-function
provides a method for determining the pion decay constant $F_{\pi}$ in
lattice gauge theory simulations.

\section{Two light flavors}

We now turn to the physically more important problem: QCD with two light
flavors.  As stated above, we follow the same steps as in the quenched case.
We first define the two-point correlation function that is very sensitive to
$F_\pi$.  This correlator can easily be obtained from a susceptibility that
we calculate in the $\epsilon$-regime of the low-energy effective theory using
the replica method.  To perform this calculation, we use a Toda lattice
equation which requires the introduction of several generating functions with
different numbers of fermionic and bosonic quarks.

We thus consider the correlation function

\be
\label{rho2Dyn}
&& \rho(\la_+,\la_-,m_u,m_d;i\mui)  \\
&&\ \ \equiv  \Bigl \langle \sum_n \delta\left(\la_+-\la_+^{(n)}\right)\sum_m
 \delta\left(\la_--\la_-^{(m)} \right)\Bigr\rangle \nn\\
&& \ \ \ \ \ \   - \Bigl\langle \sum_n \delta\left(\la_+-\la_+^{(n)} \right)\Bigr\rangle
         \Bigl\langle\sum_m \delta\left(\la_--\la_-^{(m)} \right)\Bigr\rangle\nn 
\ee
\noindent
between the eigenvalues $\la_+$ and $\la_-$ of the anti-hermitian operators
$D_+$ and $D_-$, defined in Eqs.~(\ref{D+}) and (\ref{D-}), respectively.
The average in (\ref{rho2Dyn}) is taken over the QCD partition function with
two light flavors with masses $m_u$ and $m_d$, viz.\begin{widetext}
\be
\Bigl\langle\cdots\Bigr\rangle \equiv
\frac
{\int[{\rm d}A]_\nu \ \cdots \ \det(D_++m_u)\det(D_-+m_d) e^{-S_{\rm YM}(A)}}
{\int[{\rm d}A]_\nu \det(D_++m_u)\det(D_-+m_d) e^{-S_{\rm YM}(A)}}.\nn
\ee

As in the quenched case,  the $\epsilon$-regime is reached by taking the microscopic limit of this correlator,
\be
\rho_s(\xi_+,\xi_-,\hat m_u,\hat m_d;i\hat\mui)
\equiv \lim_{V\to\infty}\frac{1}{\Sigma^2  V^2}
\rho\left(\frac{\xi_+}{\Sigma V},\frac{\xi_-}{\Sigma V},\frac{\hat{m}_u}{\Sigma
  V},\frac{\hat{m}_d}{\Sigma V};\frac{i\hatmui}{F_\pi\sqrt{V}}\right),
\ee
where $\hat m_{u,d}\equiv m_{u,d}\Sigma V$.

The scalar susceptibility,
\be
\chi(m_+,m_-,m_u,m_d;i\mui)\equiv
\left\langle \tr \frac{1}{D_++m_+}\tr
 \frac{1}{D_-+m_-}\right\rangle  
   - \left\langle \tr \frac{1}{D_++m_+}\right\rangle
         \left\langle\tr\frac{1}{D_-+m_-}\right\rangle,
\ee
which in terms of the eigenvalues $\la_+$ and $\la_-$ can be written as
\be
\chi(m_+,m_-,m_u,m_d;i\mui)  =  \left\langle \sum_n \frac{1}{i\la_+^{(n)}+m_+}\sum_m
 \frac{1}{i\la_-^{(m)}+m_-}\right\rangle  
   - \left\langle \sum_n \frac{1}{i\la_+^{(n)}+m_+}\right\rangle
         \left\langle\sum_m\frac{1}{i\la_-^{(m)}+m_-}\right\rangle,
\ee
allows us to calculate the correlation function from the discontinuity across the
imaginary axis since the QCD partition function does not depend on $m_\pm$:
\be
\rho(\lambda_+,\la_-,m_u,m_d;i\mui) & = & \frac1{4\pi^2}
{\rm  Disc}[\chi(m_+,m_-,m_u,m_d;\mui)]_{m_+=i\la_+,m_-=i\la_-} \nn\\
& = & \frac1{4\pi^2}\lim_{\ep\to0^+}[
 \chi(i\la_++\ep,i\la_-+\ep,m_u,m_d;\mui)
-\chi(i\la_+-\ep,i\la_-+\ep,m_u,m_d;\mui) \nn\\
&& \qquad\qquad -\chi(i\la_++\ep,i\la_--\ep,m_u,m_d;\mui)
+\chi(i\la_+-\ep,i\la_--\ep,m_u,m_d;\mui)] . 
\ee
The inverse of this relation is
\be
\chi(m_+,m_-,m_u,m_d;i\mui) & = & \int_{-\infty}^\infty d\la_+\, d\la_-\,
\frac{\rho(\lambda_+,\la_-,m_u,m_d;i\mui)}{(i\la_++m_+)(i\la_-+m_-)}.
\ee

\subsection{The susceptibility from the replica limit}

In order to obtain the susceptibility we again employ the replica method, writing
\be\label{chi-replica2}
\chi(m_+,m_-,m_u,m_d;i\mui) = 
\lim_{n\to0} \frac{1}{n^2}\del_{m_+}\del_{m_-}\log
{\cal Z}^\nu_{2n,2}(m_+,m_-,m_u,m_d;i\mui).
\ee
The generating functions ${\cal Z}^\nu_{2n,2}$ have 2$n$ replica flavors
in addition to the two flavors of mass $m_u$ and $m_d$,
\be
{\cal Z}^\nu_{2n,2}(m_+,m_-,m_u,m_d;i\mui)& = & \int[{\rm d}A]_\nu
\left[\det(D_++m_+)\det(D_-+m_-)\right]^n \nn \\
&& \qquad\qquad \times
 \det(D_++m_u)\det(D_-+m_d)
      e^{-S_{\rm YM}(A)}.
\label{ZnQCD}
\ee
Note that half of the replica flavors have mass $m_+$ and chemical potential
$i\mui$ while the other half have mass $m_-$ and chemical potential
$-i\mui$.

In the $\epsilon$-regime the leading contributions to the partition functions ${\cal Z}^\nu_{2n,2}$ again satisfy
Toda lattice equations. To obtain the correct replica limit $n\to0$
in Eq.~(\ref{chi-replica2}) we make use of this Toda
lattice equation \cite{SV0}:
\be
&& \hat{m}_+\del_{\hat{m}_+} \hat{m}_-\del_{\hat{m}_-} \log {\cal Z}^\nu_{2n,2}(\hat{m}_+,\hat{m}_-,\hat{m}_u,\hat{m}_d;i\hatmui)
 \\
&  &\qquad\qquad\qquad\qquad= 4 n^2(\hat{m}_+\hat{m}_-)^2\frac
  {{\cal Z}^\nu_{2n+2,2}(\hat{m}_+,\hat{m}_-,\hat{m}_u,\hat{m}_d;i\hatmui){\cal Z}^\nu_{2n-2,2}(\hat{m}_+,\hat{m}_-,\hat{m}_u,\hat{m}_d;i\hatmui)}{[{\cal Z}^\nu_{2n,2}(\hat{m}_+,\hat{m}_-,\hat{m}_u,\hat{m}_d;i\hatmui)]^2}. \nn
\ee
Taking the replica limit we arrive at
\be
\chi(\hat{m}_+,\hat{m}_-,\hat{m}_u,\hat{m}_d;i\hatmui)= 4\hat{m}_+\hat{m}_-
\frac{{\cal Z}^\nu_{2,2}(\hat{m}_+,\hat{m}_-,\hat{m}_u,\hat{m}_d;i\hatmui){\cal Z}^\nu_{2,-2}(\hat{m}_u,\hat{m}_d|\hat{m}_+,\hat{m}_-;i\hatmui)}
{[{\cal Z}^\nu_2(\hat{m}_u,\hat{m}_d;i\mui)]^2},
\label{chi-dyn}
\ee
where ${\cal Z}^\nu_{2n,2}$ has been defined in Eq.~(\ref{ZnQCD});
${\cal Z}^\nu_{2}(\hat{m}_u,\hat{m}_d;i\hatmui)
={\cal Z}^\nu_{0,2}(\hat{m}_+,\hat{m}_-,\hat{m}_u,\hat{m}_d;\hatmui)$
is the partition function with zero replica flavors; and
${\cal Z}^\nu_{2,-2}(\hat{m}_u,\hat{m}_d|\hat{m}_+,\hat{m}_-;i\hatmui)
={\cal Z}^\nu_{-2,2}(\hat{m}_+,\hat{m}_-,\hat{m}_u,\hat{m}_d;\hatmui)$.

The discontinuity of $\chi$ across the imaginary $m_+$ and $m_-$ axes
gives the dynamic correlation function,
\be
\rho_s(\xi_+,\xi_-,\hat{m}_u,\hat{m}_d;i\hatmui)= \xi_+\xi_-
\frac{{\cal Z}^\nu_{2,2}(i\xi_+,i\xi_-,\hat{m}_u,\hat{m}_d;i\hatmui)
{\cal Z}^\nu_{2,-2}(\hat{m}_u,\hat{m}_d|i\xi_+,i\xi_-;i\hatmui)}
{[{\cal Z}^\nu_2(\hat{m}_u,\hat{m}_d;i\hatmui)]^2}.
\label{2pf-dyn}
\ee
We therefore need to calculate three different generating functions: two  with fermionic quarks only, and one with both fermionic and bosonic quarks.

\subsection{Computing the fermionic partition function}

In the effective theory in the $\epsilon$-regime, the generating functions
with $n\geq0$ involved in the calculation of the susceptibility are given by
\be
{\cal Z}^\nu_{2n,2}(\hat{m}_+,\hat{m}_-,\hat{m}_u,\hat{m}_d;i\hatmui) =
\int_{U \in U(2n+2)} dU\, \det(U)^\nu e^{\frac{1}{4}\hatmui^2
\Tr [U,B][U^\dagger,B] + \frac 12 \Tr(M^\dagger U + MU^\dagger)},
\label{zeff}
\ee
with
\be
B = {\rm diag}({\mathbf 1}_{n+1},-{\mathbf 1}_{n+1}) \qquad {\rm and} \qquad M
= {\rm diag}(\hat{m}_+,...,\hat{m}_+,\hat{m}_u,\hat{m}_-,...,\hat{m}_-,\hat{m}_d).
\ee
The partition function ${\cal Z}^\nu_2(\hat{m}_u,\hat{m}_d;i\hatmui)$ has
been calculated in Eq.~(\ref{zeff2mui}) above. In addition, ${\cal
  Z}^\nu_{2,2}(\hat{m}_+,\hat{m}_-,\hat{m}_u,\hat{m}_d;i\hatmui)$ can be obtained from the
calculation presented in Ref.~\cite{AFV} by changing the sign of $\hatmui^2$. We have thus
\be
\label{Zeff22}
{\cal Z}^\nu_{2,2}(\hat{m}_+,\hat{m}_-,\hat{m}_u,\hat{m}_d;i\hatmui) =
\frac{1}{(\hat{m}_+^2-\hat{m}_u^2)(\hat{m}_-^2-\hat{m}_d^2)}\left|\begin{array}{cc}
{\cal Z}^\nu_2(\hat{m}_+,\hat{m}_-;i\hatmui) &
{\cal Z}^\nu_2(\hat{m}_+,\hat{m}_d;i\hatmui)\\[2pt]
{\cal Z}^\nu_2(\hat{m}_u,\hat{m}_-;i\hatmui) &
{\cal Z}^\nu_2(\hat{m}_u,\hat{m}_d;i\hatmui)
\end{array}\right|.
\ee

\subsection{Computing the supersymmetric partition functions}
The calculation of ${\cal
  Z}^\nu_{2,-2}(\hat{m}_u,\hat{m}_d|\hat{m}_+,\hat{m}_-;i\hatmui)$ requires
us to perform an exact integral over the supergroup $\hat{\rm Gl}(2|2)$ with two
fermionic and two bosonic quark flavors. This is a rather lengthy analytical
calculation.
We start from the effective generating function in the $\epsilon$-regime which is given by
\begin{eqnarray}
  \label{zeroMomPF}
  {\cal Z}^\nu_{2,-2}(\hat{m}_u,\hat{m}_d|\hat{m}_+,\hat{m}_-;i\hatmui)
=\int_{\hat{\rm Gl}(2|2)} dU\, {\rm Sdet}^\nu U e^{
\frac{1}4 \hatmui^2  \Str [B,U] [B,U^{-1}]
+\frac{1}2 \Str {\cal M} (U+U^{-1})},
\end{eqnarray}
\end{widetext}
where $B={\rm diag}(1,1,-1,-1)$, and the mass matrix is given by
\begin{eqnarray}
  {\cal M} = \mat M_+ & 0 \\ 0 & M_- \emat,
\end{eqnarray}
with $M_+={\rm diag}(\hat{m}_u,\hat{m}_+)$ and $M_-={\rm
  diag}(\hat{m}_d,\hat{m}_-)$ \cite{TVmuRealQCD}. At $\mui=0$,
this is the same as the generating function used in \cite{TV}. Notice that in
the complete effective partition function
at $\mui=0$ there is an invariant operator that contains the term $\partial_\nu
\Str \ln U=\Str U^{-1} \partial_\nu U$ \cite{OTV,DOTV}. This operator is obviously
absent from the zero-momentum part of the partition function. At
non-zero $\mui$, this operator is also absent from the zero-momentum
part of the partition function since \cite{TVmuRealQCD} $\Str U^{-1} \nabla_0 U=\Str
U^{-1} \partial_0 U$, and $\nabla_0 U=\partial_0 U-i\mui [B,U]$ . We can understand this from physics as well: The isospin singlet does not couple to isospin chemical potential.

\subsubsection{Parameterization of the Goldstone supermanifold}
In order to calculate the exact supergroup integral (\ref{zeroMomPF}),
we have to parameterize the Goldstone manifold.  We use the same factorizing
parameterization as in Ref.~\cite{TV}:
\begin{equation}
  \label{paramU}
  U=\mat w_1 & 0 \\ 0 & w_2 \emat
       \mat \sqrt{1-w\bar{w}} & w \\ -\bar{w} & \sqrt{1-\bar{w}w}\emat
       \mat w_1 & 0 \\ 0 & w_2 \emat,
\end{equation}
where $w_{1,2}\in \hat{\rm Gl}(1|1)$ and $w, \; \bar{w} \in {\rm
  Gl}(1|1)$. This parameterization leads to rather simple
expressions for the integrand of (\ref{zeroMomPF}):
\begin{eqnarray}
\label{Sdet}
{\rm Sdet} U={\rm Sdet} (w_1^2 \; w_2^2),
\end{eqnarray}
\begin{eqnarray}
  \label{zeroMomAction}
  \frac{1}2 \Str {\cal M} (U+U^{-1}) \equiv {\cal S}_1 + {\cal S}_2,
\end{eqnarray}
and
\begin{eqnarray}
\frac{1}4 \hatmui^2 \Str [B,U][B,U^{-1}] \equiv {\cal S}_\mu,
\end{eqnarray}
with
\begin{eqnarray}
  \label{S1}
  {\cal S}_1&=&\frac12 \Str \left[ M_+ \left( w_1 \sqrt{1-w\bar{w}}\, w_1\right.\right.\nn\\
  &&\qquad\quad\left.\left.+
      w_1^{-1} \sqrt{1-w\bar{w}}\, w_1^{-1}  \right) \right],
\end{eqnarray}
\begin{eqnarray}
  \label{S2}
  {\cal S}_2&=&\frac12 \Str \left[ M_- \left( w_2 \sqrt{1-\bar{w}w} \,w_2
  \right.\right.\nn\\
  &&\qquad\quad\left.\left.+
      w_2^{-1} \sqrt{1-\bar{w}w} \,w_2^{-1}  \right) \right],
\end{eqnarray}
and
\begin{eqnarray}
  \label{Smu}
  {\cal S}_\mu=-2 \hatmui^2  \Str  \left[ w \bar{w}  \right].
\end{eqnarray}
For $w^2_{1,2} \in \hat{\rm Gl}(1|1)$, we use the same
parameterization as the one used in Ref.~\cite{DOTV},
\begin{eqnarray}
  \label{wi}
  w^2_i=\Lambda_i v_i ,
\end{eqnarray}
where
\begin{eqnarray}
\Lambda_i=\mat e^{i \psi_i} & 0 \\ 0 & e^{s_i} \emat   \quad
{\rm and} \quad
v_i=\exp \mat 0 & \alpha_i \\ \beta_i & 0  \emat,
\end{eqnarray}
with $\psi_i\in(-\pi,\pi)$ and $s_i\in(-\infty,\infty)$;
$\alpha_i$ and $\beta_i$ are Grassmann variables.
The main advantage of this parameterization is that its
Berezinian is equal to $1$, as was shown in Ref.~\cite{DOTV}. Thus we parameterize the matrices $w_{1,2} \in \hat{\rm
Gl}(1|1)$ so that $w_i^2=\Lambda_i v_i$, i.e.,
\begin{eqnarray}
w_i=\mat  (1+\frac12 c_i \alpha_i \beta_i)\;e^{i \psi_i/2} & a_i \alpha_i \\
b_i \beta_i &   (1-\frac12 d_i \alpha_i \beta_i)\;e^{s/2}  \emat,
\end{eqnarray}
where
\begin{eqnarray}
a_i&=&\frac{e^{i \psi_i}}{e^{i\psi_i/2}+e^{s_i/2}} \nn \\
b_i&=&\frac{e^{s_i}}{e^{i\psi_i/2}+e^{s_i/2}} \\
c_i&=&\frac12 - \frac{e^{s_i}}{(e^{i\psi_i/2}+e^{s_i/2})^2} \nn \\
d_i&=&\frac12 - \frac{e^{i \psi_i}}{(e^{i\psi_i/2}+e^{s_i/2})^2}. \nn
\end{eqnarray}
For $w, \, \bar{w} \in {\rm Gl}(1|1)$ we use the same polar
decomposition as in \cite{TV}:
\begin{eqnarray}
  \label{w}
  w=vSu^{-1} \quad {\rm and} \quad \bar{w}=u\bar{S}v^{-1},
\end{eqnarray}
where $S$ and $\bar{S}$ are $2\times2$ diagonal supermatrices with
commuting elements given by
\begin{eqnarray}
  S&=&{\rm diag}(\sin \theta e^{i\rho},i \sinh \phi e^{i\sigma})
  ,\\ \bar{S}&=&{\rm diag}(\sin \theta
  e^{-i\rho},i \sinh \phi e^{-i\sigma}),
\end{eqnarray}
with $\theta\in(0,\pi/2)$, $\phi\in(0,\infty)$, and
$\rho,\sigma\in(-\pi,\pi)$, while $u, \, v \in U(1|1)/[U(1)\times U(1)]$
are given by \cite{TV}
\begin{eqnarray}
  u=\exp \mat 0 & \zeta \\ \chi & 0 \emat \quad {\rm and}
  \quad v=\exp \mat 0 & \xi \\ \eta & 0 \emat.
\end{eqnarray}

With these parameterizations, we obtain
\begin{eqnarray}
\label{Sdetdiag}
{\rm Sdet}\,U=\prod_i e^{i\psi_i-s_i}.
\end{eqnarray}
In addition, the supertraces in Eqs.~(\ref{S1}),
(\ref{S2}), and~(\ref{Smu}) are given by
\begin{equation}
  \label{S1diag}
{\cal S}_1=\frac12 \Str\left[ vCv^{-1} \left( w_1 M_+
    w_1 +  w_1^{-1} M_+ w_1^{-1}  \right)  \right],
\end{equation}
\begin{equation}
  \label{S2diag}
{\cal S}_2=\frac12 \Str\left[ uCu^{-1} \left( w_2 M_-
    w_2 +  w_2^{-1}  M_- w_2^{-1}  \right)  \right],
\end{equation}
and
\begin{eqnarray}
  \label{Smudiag}
  {\cal S}_\mu=2\hatmui^2 \Str C^2,
\end{eqnarray}
where $C=\sqrt{1-S\bar{S}}={\rm diag}(\cos\theta,\cosh\phi)$.
The advantage of our parameterization is that the integral over the
supergroup $\hat{\rm Gl}(2|2)$ explicitly contains two independent integrals over
$\hat{\rm Gl}(1|1)$, which are simpler to compute analytically.

In order to perform the group integral we need to determine the integration
measure that corresponds to our parameterization.
The parameterization of the Goldstone manifold (\ref{paramU}) is of
the form
\begin{eqnarray}
  U=W T W.
\end{eqnarray}
As was shown in Ref.~\cite{TV}, the measure factorizes into a product
of one factor that depends only on $W$ and one factor that depends
only on $T$,
\begin{eqnarray}
  d\mu(U)&=& w_1^{-2} \,dw_1^2\, w_2^{-2}\, dw_2^2\, T^{-1}\,dT \nn \\
&\equiv&\mu(w_1)\, dw_1\, \mu(w_2)\,dw_2\,\mu(w,\bar{w}) dw\, d\bar{w}.
\end{eqnarray}
For $w^2_{1,2} \in \hat{\rm Gl}(1|1)$, we have used the same
parameterization as in Ref.~\cite{DOTV}. In that paper, it was shown
that the Berezinian of this change of variables is equal to $1$, and thus that
\begin{eqnarray}
  w_i^{-2}\, dw_i^2=d\psi_i\, ds_i\, d\alpha_i\, d\beta_i.
\end{eqnarray}
We therefore find that
\begin{eqnarray}
  \label{mu(wi)}
\mu(w_i) \,dw_i=ds_i\, d\psi_i \,d\alpha_i\, d\beta_i.
\end{eqnarray}
Finally, the parameterization we use for $T$ is exactly the same as the one
used in Ref.~\cite{TV}. The measure is given by
\begin{eqnarray}
  T^{-1}\,dT&=&\mu(w,\bar{w})\, dw\, d\bar{w}\nn\\
  &=&\frac{i \sinh 2\phi \sin
    2\theta}{(\cos^2 \theta-\cosh^2 \phi)^2} d\theta\, d\phi\,d\rho
\,  d\sigma \,d\zeta\, d\chi\, d\xi \,d\eta.\nn\\
  \label{muw}
\end{eqnarray}

\subsubsection{Efetov-Wegner terms}

As for any supersymmetric integral, extreme care has to be taken with the
singularities that might be introduced through a specific
parameterization of the integration supermanifold and the corresponding
measure. The singularities of the measure affect the supersymmetric
integral through the so called Efetov-Wegner terms. (See for example
Refs.~\cite{GPW,DOTV} for a discussion of Efetov-Wegner terms.) 
The measure $\mu(w_i)\, dw_i$ does not contain any singularity in the
variables $s_i$ and $\psi_i$, and there are no Efetov-Wegner
terms related to our parameterization of $w_{1,2}$. 
On the other hand, the
measure $T^{-1}\,dT=\mu(w,\bar{w})\, dw\, d\bar{w}$ is singular when $\theta=i
\phi$. We therefore expect Efetov-Wegner terms in this case.  The method used
in \cite{GPW} can be straightforwardly applied
to compute the Efetov-Wegner terms related to our parameterization of $T$.
With our parameterization, including the Efetov-Wegner terms,  we find that
the generating function ${\cal Z}^\nu_{2,-2}$ is given by 
\begin{widetext}
\begin{eqnarray}
  \label{partFct}
  {\cal Z}^\nu_{2,-2}(\hat{m}_u,\hat{m}_d|\hat{m}_+,\hat{m}_-;i\hatmui)&=&\frac1{(2\pi)^3}\lim_{\epsilon\rightarrow0}\int
  d\theta \,d\phi\, d\rho\, d\sigma\, d\zeta\, d\chi\, d\xi\, d\eta\, \frac{i \sinh 2\phi \sin
    2\theta}{(\cos^2 \theta-\cosh^2 \phi)^2} \\
&&\hspace{0.5cm} \times\left[ \theta(v_c-\epsilon)+
    \delta(v_c-\epsilon) v_n+\frac12 \delta'(v_c-\epsilon) v_n^2
  \right]  {\cal I}_1 {\cal I}_2 e^{ \hatmui^2 (\cos^2\theta-\cosh^2
    \phi)}, \nn
\end{eqnarray}
where
\begin{eqnarray}
  v_c&=&\sin^2 \theta+\sinh^2 \phi \nn \\
 v_n&=&(\sin^2 \theta-\sinh^2 \phi) (\zeta \chi+\xi \eta)
+2i \sin\theta \sinh\phi \left( e^{i(\rho-\sigma)} \eta\zeta -
   e^{-i(\rho-\sigma)} \xi\chi \right) \nn \\
&&+2 (\sin^2 \theta+\sinh^2 \phi )  \zeta\chi\xi\eta \\
v_n^2&=&2 (\sin^2 \theta+\sinh^2 \phi )^2 \zeta\chi\xi\eta, \nn
\end{eqnarray}
and
\begin{eqnarray}
  \label{Ii}
  {\cal I}_i=\frac1{2\pi}\int_{-\pi}^\pi d\psi_i \int_{-\infty}^\infty ds_i
  \int d\alpha_i\, d\beta_i \, e^{{\cal S}_i+\nu (i\psi_i-s_i)}.
\end{eqnarray}
The normalization is chosen so that $Z^\nu_{2,-2}(\hat{m}_u,\hat{m}_d|\hat{m}_u,\hat{m}_d;i\hatmui)=1$.

\subsubsection{Analytical result for the supersymmetric partition function}

We are now in position to explicitly compute the partition function (\ref{zeroMomPF}).
We first compute ${\cal I}_i$ given by Eq.~(\ref{Ii}). The integral over
$\alpha_i$ and $\beta_i$ is readily obtained by expanding $\exp {\cal
  S}_i$ to first order in $\alpha_i$ and $\beta_i$.
We get
\begin{eqnarray}
\label{Iint}
  {\cal I}_1&=&{\cal J}(\hat{m}_u,\hat{m}_+,\cos \theta, \cosh \phi) + \xi\eta
  {\cal K}(\hat{m}_u,\hat{m}_+,\cos \theta, \cosh \phi),  \nonumber \\[2pt]
{\cal I}_2&=&{\cal J}(\hat{m}_d,\hat{m}_-,\cos \theta,\cosh \phi)+\zeta\chi
{\cal   K}(\hat{m}_d,\hat{m}_-,\cos \theta,\cosh \phi),
\end{eqnarray}
with
\begin{eqnarray}
{\cal J}(x,y,t,p)&=& \frac1{16\pi} \int d\psi ds \; e^{tx \cos\psi-py\cosh s} \frac{e^{\nu (i\psi-s)}}{1+\cosh\frac{s-i\psi}2} \nonumber\\[2pt]
&& \qquad\quad \times \left[ 2tx\cos\psi+2py\cosh s+py \cosh\frac{3s-i\psi}2+ (2px-tx-py+2ty)\cosh\frac{s+i\psi}2 \right.\\
&&\qquad\qquad\quad\left.+tx \cosh\frac{s-3i\psi}2 \right], \nonumber
\end{eqnarray}
and
\begin{eqnarray}
{\cal K}(x,y,t,p)&=&\frac{t-p}{64\pi} \int d\psi ds \; e^{tx \cos\psi-py\cosh s} \;e^{\nu (i\psi-s)} \nonumber\\
&& \qquad\quad\times  \big[ px^2-ty^2 + x(2\cos\psi+tx \cos2\psi) + y (2\cosh s-py \cosh 2s) \\
&& \qquad\qquad\quad  +2 i (t-p) xy \sin\psi \sinh s \big]. \nonumber
\end{eqnarray}
Hence the partition function (\ref{partFct}) can be written as
\begin{eqnarray}
  \label{parFct2}
Z^\nu_{2,-2}(\hat{m}_u,\hat{m}_d|\hat{m}_+,\hat{m}_-;i\hatmui)&=& \lim_{\epsilon\rightarrow0} \int_0^1 dt
\int_1^\infty dp \frac{tp}{(t^2-p^2)^2}
e^{2\hatmui^2 (t^2-p^2)} \theta(p^2-t^2-\epsilon) \nn \\[2pt]
 && \qquad\qquad\quad\times{\cal K}(\hat{m}_u,\hat{m}_+,t,p) {\cal K}(\hat{m}_d,\hat{m}_-,t,p) \nn \\
&&+\lim_{\epsilon\rightarrow0} \int_0^1 du
\int_0^\infty dv
\frac{uv}{(u^2+v^2)^2}
e^{\hatmui^2 (u^2+v^2)}
\delta(u^2+v^2-\epsilon) \\
&&\quad\times\left\{ (u^2-v^2) \left[ {\cal
  K}\left(\hat{m}_u,\hat{m}_+,\sqrt{1-u^2},\sqrt{1+v^2}\right)
{\cal  J}\left(\hat{m}_d,\hat{m}_-,\sqrt{1-u^2},\sqrt{1+v^2}\right) \right.\right.\nn \\
&&\left.\qquad\qquad+{\cal J}\left(\hat{m}_u,\hat{m}_+,\sqrt{1-u^2},\sqrt{1+v^2}\right)
{\cal  K}\left(\hat{m}_d,\hat{m}_-,\sqrt{1-u^2},\sqrt{1+v^2}\right)     \right]  \nn \\
&&\qquad\left.+(u^2+v^2) {\cal
  J}\left(\hat{m}_u,\hat{m}_+,\sqrt{1-u^2},\sqrt{1+v^2}\right) {\cal
  J}\left(\hat{m}_d,\hat{m}_-,\sqrt{1-u^2},\sqrt{1+v^2}\right) \right\}, \nn
\end{eqnarray}
where $t=\cos\theta$, $p=\cosh\phi$, $u=\sin\theta$, and $v=\sinh\phi$.
We can rewrite the second integral as
\begin{eqnarray}
&&\hspace{-7mm}\lim_{\epsilon\rightarrow0} 
\int_0^{\pi/2} dq \int_0^{1/\sin q} dr \,\sin2q  \, e^{-2\hatmui^2 r}
\delta(r-\epsilon) \Bigg\{ (\cos^2q-\sin^2q) \nn \\
&&\times\left[ {\cal
  K}\left(\hat{m}_u,\hat{m}_+,\sqrt{1+r \cos^2q},\sqrt{1-r \sin^2q}\right)
{\cal  J}\left(\hat{m}_d,\hat{m}_-,\sqrt{1+r \cos^2q},\sqrt{1-r \sin^2q}\right)\right. \nn \\
&&\quad\left.+{\cal J}\left(\hat{m}_u,\hat{m}_+,\sqrt{1+r \cos^2q},\sqrt{1-r \sin^2q}\right)
{\cal  K}\left(\hat{m}_d,\hat{m}_-,\sqrt{1+r \cos^2q},\sqrt{1-r \sin^2q}\right)     \right]  \nn \\
&&+ {\cal
  J}\left(\hat{m}_u,\hat{m}_+,\sqrt{1+r \cos^2q},\sqrt{1-r \sin^2q}\right) {\cal
  J}\left(\hat{m}_d,\hat{m}_-,\sqrt{1+r \cos^2q},\sqrt{1-r \sin^2q}\right) \Bigg\} \nn \\
&=& {\cal J}(\hat{m}_u,\hat{m}_+,1,1) {\cal J}(\hat{m}_d,\hat{m}_-,1,1),
\end{eqnarray}
where $u=\sqrt{r} \cos q$ and $v=\sqrt{r} \sin q$.
This gives
\begin{eqnarray}
{\cal Z}^\nu_{2,-2}(\hat{m}_u,\hat{m}_d|\hat{m}_+,\hat{m}_-;i\hatmui)&=&\int_0^1 dt
\int_1^\infty dp \,\frac{tp}{(t^2-p^2)^2}  e^{2\hatmui^2 (t^2-p^2)} {\cal K}(\hat{m}_u,\hat{m}_+,t,p) {\cal K}(\hat{m}_d,\hat{m}_-,t,p) \nonumber\\[2pt]
&& +{\cal J}(\hat{m}_u,\hat{m}_+,1,1) {\cal J}(\hat{m}_d,\hat{m}_-,1,1).
\end{eqnarray}
Finally, we have to calculate ${\cal J}(x,y,1,1)$ and ${\cal K}(x,y,t,p)$. The result is
\begin{eqnarray}
{\cal J}(x,y,1,1)=\frac12 \left\{ x \left[ I_{\nu-1}(x)+I_{\nu+1}(x) \right] K_\nu(y)+ y I_\nu(x) \left[ K_{\nu-1}(y)+K_{\nu+1}(y) \right] \right\},
\end{eqnarray}
and
\begin{eqnarray}
{\cal K}(x,y,t,p)=(x^2-y^2) (t^2-p^2) I_\nu(tx) K_\nu(py).
\end{eqnarray}

The partition function can finally be written as
\begin{eqnarray}
{\cal Z}^\nu_{2,-2}(\hat{m}_u,\hat{m}_d|\hat{m}_+,\hat{m}_-;i\hatmui)&=&\left[(\hat{m}_u)^2- (\hat{m}_+)^2\right]
    \left[(\hat{m}_d)^2- (\hat{m}_-)^2\right] \nonumber\\
    && \hspace{-1cm} \times\int_0^1 dt \, t  e^{2\hatmui^2 t} \, I_\nu(t\hat{m}_u) I_\nu(t\hat{m}_d)
     \int_1^\infty dp \, p  e^{-2\hatmui^2 p} \, K_\nu(p\hat{m}_+) K_\nu(p\hat{m}_-) \\
   &&\hspace{-1cm} + \frac14
\left\{\hat{m}_u \left[ I_{\nu-1}(\hat{m}_u)+I_{\nu+1}(\hat{m}_u) \right] K_\nu(\hat{m}_+)+ \hat{m}_+
I_\nu(\hat{m}_u) \left[ K_{\nu-1}(\hat{m}_+)+K_{\nu+1}(\hat{m}_+) \right] \right\}\nn\\
&&\hspace{-1cm}\quad\times\left\{ \hat{m}_d \left[ I_{\nu-1}(\hat{m}_d)+I_{\nu+1}(\hat{m}_d) \right]
K_\nu(\hat{m}_-)+ \hat{m}_- I_\nu(\hat{m}_d) \left[ K_{\nu-1}(\hat{m}_-)+K_{\nu+1}(\hat{m}_-) \right] \right\}. \nonumber
\end{eqnarray}
This can be written in a more compact notation as
[cf.~Eqs.~(\ref{zeff2mui}) and~(\ref{Z-1mui})]
\be\label{Z2m2}
{\cal Z}^\nu_{2,-2}(\hat{m}_u,\hat{m}_d|\hat{m}_+,\hat{m}_-;i\hatmui) =
\left|\begin{array}{cc}
(\hat{m}_u^2-\hat{m}_+^2){\cal Z}^\nu_2(\hat{m}_u,\hat{m}_d;i\hatmui) &  {\cal Z}^\nu_{1,-1}(\hat{m}_d|\hat{m}_-)\\[2pt]
-{\cal Z}^\nu_{1,-1}(\hat{m}_u|\hat{m}_+) &  (\hat{m}_d^2-\hat{m}_-^2){\cal Z}^\nu_{-2}(\hat{m}_+,\hat{m}_-;i\hatmui)
\end{array}\right|,
\ee
where the $\mui$-independent graded partition function,
\be
{\cal Z}^\nu_{1,-1}(\hat{m}_u|\hat{m}_+) =
\frac12
\left\{ \hat{m}_u \left[ I_{\nu-1}(\hat{m}_u)+I_{\nu+1}(\hat{m}_u) \right] K_\nu(\hat{m}_+)+ \hat{m}_+
I_\nu(\hat{m}_u) \left[ K_{\nu-1}(\hat{m}_+)+K_{\nu+1}(\hat{m}_+) \right] \right\},
\label{Zsusy}
\ee
was calculated in Ref.~\cite{DOTV}.

Note that ${\cal Z}^\nu_{2,-2}(\hat{m}_+,\hat{m}_-|\hat{m}_+,\hat{m}_-;i\hatmui)=1$ as it should.
Furthermore, upon expanding ${\cal Z}^\nu_{2,-2}(\hat{m}_u,\hat{m}_d|\hat{m}_+,\hat{m}_-;i\hatmui)$ to leading
order in $m_u-m_+$ and $m_d-m_-$, the quenched correlation function (\ref{rho2Qmui})
is recovered using the supersymmetric method.

\subsection{Final result}
We can thus finally compute the two-point correlation function
(\ref{2pf-dyn}), using the analytically calculated generating functions
(\ref{zeff2mui}), (\ref{Zeff22}), and (\ref{Z2m2}).
The result is
\be\label{rho-result}
\rho_s(\xi_+,\xi_-,\hat{m}_u,\hat{m}_d;i\hatmui)
& = & \xi_+\xi_- \left[
\int_0^1 dt \;t \;e^{2\hatmui^2t^2} I_\nu(t \;\hat{m}_u)
I_\nu (t \;\hat{m}_d)\right]^{-2} \\
& & \times \left(
\int_0^1 dt \;t \;e^{2\hatmui^2t^2} J_\nu(t \;\xi_+)
J_\nu (t \;\xi_-) \int_0^1 dt \;t \;e^{2\hatmui^2t^2}
I_\nu(t \;\hat{m}_u)I_\nu (t \;\hat{m}_d) \right.\nn\\
& & -\left.
\int_0^1 dt \;t \;e^{2\hatmui^2t^2} I_\nu(t \;\hat{m}_u)
J_\nu (t \;\xi_-) \int_0^1 dt \;t\;
e^{2\hatmui^2t^2} J_\nu (t \;\xi_+) I_\nu(t \;\hat{m}_d) \right)\nn\\
&& \times\left( \int_0^1 dt \;t \;e^{2\hatmui^2t^2} I_\nu(t \;\hat{m}_u)
I_\nu (t \;\hat{m}_d)\int_1^\infty dt \;t \;e^{-2\hatmui^2t^2}
J_\nu(t \;\xi_+)J_\nu (t \;\xi_-)\right. \nn\\
&&\hspace{-2cm}\left.+\frac{(\hat{m}_u I_{\nu+1}(\hat{m}_u) J_\nu(\xi_+)+\xi_+ J_{\nu+1}(\xi_+)I_\nu(\hat{m}_u) )(\hat{m}_d I_{\nu+1}(\hat{m}_d) J_\nu(\xi_-)+\xi_- J_{\nu+1}(\xi_-)I_\nu(\hat{m}_d) )}{(\xi_+^2+\hat{m}_u^2)(\xi_-^2+\hat{m}_d^2)}\right).\nn
\ee
\end{widetext}
For a numerical evaluation it is advantageous to rewrite the non-compact
integral appearing in the fourth line as in Eq.~(\ref{rho2Qmui}).

We have performed various checks on this result. For example,
at $\hatmui = 0$ it correctly reduces to the two-point microscopic
correlation functions at zero chemical potential \cite{OTV,DOTV}. We have
also verified that it reduces to the quenched result (\ref{rho2Qmui})
in the limit where both $m_u$ and $m_d$ are sent to infinity, as
is required by decoupling.

\begin{figure}[htb]
\begin{center}
\includegraphics[width=.9\columnwidth]{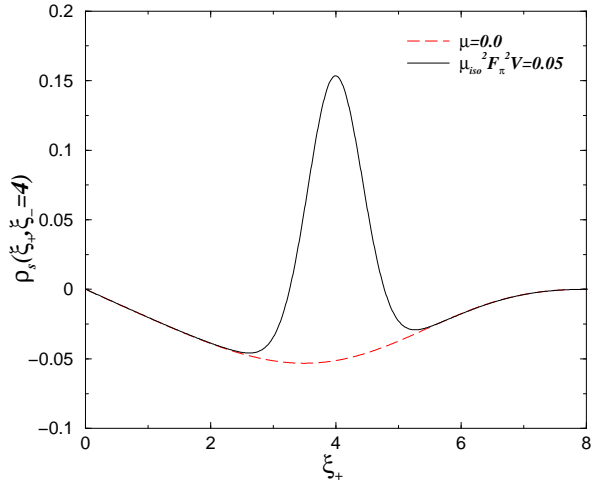}
\caption{\label{fig:2}
The two-point correlation function (\ref{rho-result}), with one eigenvalue
fixed at $\xi_- = 4$, as a function of the other eigenvalue $\xi_+$. The
masses of the two dynamical flavors are chosen to be degenerate at the value
$\hat{m}_u=\hat{m}_d=5$. The dashed curve shows the correlation function for 
$\mui=0$ and the solid curve corresponds to $\mui^2F_{\pi}^2V = 0.05$.
The $\delta$-function that appears at $\xi_+=\xi_-=4$ when $\mui=0$ has been
suppressed from this figure.}
\end{center}
\end{figure}

As in the quenched case the correlation function at $\mui=0$ has a $\delta$-function 
at equal arguments. When $\mui$ is nonzero this $\delta$-function
becomes a peak in the correlation function around equal arguments, as shown in
Fig.~\ref{fig:2}.

\section{Conclusion}

We have considered QCD at nonzero imaginary isospin chemical
potential and made use of its effective field theory representation to calculate a
correlation function between eigenvalues of the Dirac operator that is
very sensitive to the value of the pion decay constant $F_\pi$.  We have
shown the calculation in the quenched case as well as in the physical
case with two dynamical light quark flavors. In two previous articles, it
has been demonstrated
that these formulas for the correlation functions lead to an efficient way to
determine $F_\pi$ on the lattice \cite{DHSS,DHSST}.

Our calculation made extensive use of the replica method.  In this
approach, the correct answer is reached via Toda lattice equations,
which relate theories with differing numbers of quark species to each
other. Taking the replica limit required the computation of several partition functions with varying
numbers of bosonic and fermionic quarks, which led us to compute
some non-trivial partition functions.  Our
strategy thus required the calculation of exact group integrals over
both graded and non-graded Goldstone manifolds.  It must be noted
that our calculation would be technically very tedious if it were carried out exclusively by
the so-called supersymmetric approach:  It would require an integration over
$\hat{\rm Gl}(4|2)$, a highly complicated task.  The advantage of the Toda lattice equation is that it
reduces the complexity of the calculation by spreading the difficulty over
several partition functions.

As a tool to extract $F_\pi$ from dynamical lattice simulations it
is of obvious interest to derive the correlation function
in a partially quenched theory where the chemical potential is set to zero for the physical dynamical quarks. This would allow the use of existing gauge field
ensembles, generated at zero chemical potential, in a determination of $F_\pi$.
The analytical expression for this partially quenched correlation function
has proved to be challenging. We hope that the detailed calculation presented
here may be of help in the future calculation of such quantities.

\vspace{1cm}

\section*{Acknowledgements}
We thank G.~Akemann and J.~Verbaarschot for useful discussions.
The work of BS was supported in part by the Israel Science
Foundation under grant no.~173/05.  He thanks the Niels Bohr
Institute for its hospitality.
The work of DT was supported by the NSF under grant
no.~NSF-PHY0304252. He thanks the Particle Physics Group at the
Rensselaer Polytechnic Institute for its hospitality. KS was supported by the
Carlsberg Foundation.

\end{document}